\documentclass[prl,aps,twocolumn,superscriptaddress,preprintnumbers,showpacs]{revtex4}

\usepackage{amsmath,amsfonts,amssymb,graphicx,color,times,bbm,psfrag,dsfont,
slashed}
\usepackage[colorlinks=true,citecolor=blue]{hyperref}

\newcommand{\be}{\begin{equation}}
\newcommand{\ee}{\end{equation}}

\usepackage[T1]{fontenc}
\usepackage[latin9]{inputenc}

\newcommand{\bra}[1] {\langle{#1}\vert}
\newcommand{\ket}[1] {\vert{#1}\rangle}
\newcommand{\beq}{\begin{equation}}
\newcommand{\eeq}{\end{equation}}
\newcommand{\bea}{\begin{eqnarray}}
\newcommand{\eea}{\end{eqnarray}}

\renewcommand{\b}[1]{\mathbf{ #1}}									

\begin{document}

\title{Edge insulating topological phases in a two-dimensional  superconductor with long-range pairing}

\author{L. Lepori}
\email[correspondence at: ]{llepori81@gmail.com}
\affiliation{Dipartimento di Scienze Fisiche e Chimiche, Universit\`a dell'Aquila, via Vetoio,
I-67010 Coppito-L'Aquila, Italy.}
\affiliation{INFN, Laboratori Nazionali del Gran Sasso, Via G. Acitelli,
22, I-67100 Assergi (AQ), Italy.}

\author{D. Giuliano}
\affiliation{Dipartimento di Fisica, Universit\`a della Calabria, I-87036 Arcavacata di Rende,  Cosenza, Italy.}
\affiliation{INFN, Gruppo collegato di Cosenza, I-87036 Arcavacata di Rende, Cosenza, Italy.}

\author{S. Paganelli}
\affiliation{Dipartimento di Scienze Fisiche e Chimiche, Universit\`a dell'Aquila, via Vetoio,
I-67010 Coppito-L'Aquila, Italy.}

\begin{abstract}
We study the zero-temperature phase diagram of a two dimensional square lattice  loaded by spinless fermions, with nearest-neighbor hopping and algebraically decaying  pairing. We find that for sufficiently long-range pairing, {\color{black} new phases occur, not continuously connected with any short-range phase and not belonging to the standard families of topological insulators/superconductors.  These phases are} signaled by the violation of the area law for the Von Neumann entropy, by semi-integer Chern numbers, and by edge modes with nonzero mass. The latter feature results in the absence of single-fermion edge conductivity,  present instead in the short-range limit. The definition of a bulk-topology and the presence of a bulk-boundary correspondence is suggested also for the long-range phases. Recent experimental proposals and advances open the possibility to probe the described long-range effects in near-future realistic set-ups.
\end{abstract}

\maketitle

{\bf Introduction --}
In recent years, the topological phases of matter
have become  a central focus for  physical investigation.  
A groundbreaking result is the 
classification of the symmetry-protected topologically inequivalent {\color{black} classes for non interacting fermionic systems
(topological insulators/superconductors)}
\cite{zirnbauer1996,zirnbauer1997,ludwig2008,kitaev2009,ludwig2009,ludwig2010}.
This theoretical breakthrough has been probed  
 on particular solid-state 
 compounds \cite{hasankane,zhang11,bernevigbook,chiu15}. 

Notwithstanding the presence of a nonvanishing bulk energy gap, 
the most relevant feature
displayed by  nontrivial topological insulators/superconductors
is a conductivity localized on the edges, due to massless edge mode with a dynamics well distinguished
from the bulk excitations. Moreover, continuous transitions with a vanishing mass gap generally divide phases with different topology (even if also first order transitions seem possible if perturbative interactions between fermions are added \cite{capone2015}).
Regarding the entanglement content, 
{\color{black} topological insulators/superconductors} display exponentially saturating entanglement and correlations, giving rise to the area-law for the Von Neumann
entropy between the two elements of a real-space bipartition.

These  features characterize quantum systems governed by Hamiltonians
with short-range (SR) terms.  However, in recent years,   
long-range (LR) classical and quantum systems \cite{libro}, obtained 
renewed attention.
Independent theoretical works have concluded that LR systems 
can exhibit many interesting and unusual properties, essentially due to the 
violation of  locality 
(see e.g. \cite{deng2005,hast,koffel2012,hauke2013,nostro,paperdouble,maghrebi2015-2,santos2015}). 
In particular,  \emph{one-dimensional} LR models can host new phases, manifesting striking properties,  not present in the SR limit.
\cite{nostro,ares,paperdouble,delgado2015,
paper1,gori,gong2015,maghrebi2015-2,gong2015-2,lepori2016mech,lepori2016,alecce2017,giuliano2017}.

In spite of these relevant achievements,
full comprehension and classification {\color{black} phases emerging from Hamiltonians with LR terms} 
are still not available.

In \cite{lepori2016} a partial solution to these problems has been discussed, exploiting one-dimensional topological superconductive chains as playgrounds, but
inferring nontrivial properties also for LR topological insulators/superconductors in higher dimensions. 

From the experimental side,  recent proposals with Shiba states \cite{oppen2013,menard2015,ojanen2015,bernevig2016}, Floquet Hamiltonians \cite{platero2015,you2017},  atoms coupled to radiation fields \cite{bettles2017,perczel2017,perczel2017-bis}, and trapped ions \cite{porras2017} opened the possibility to synthesize 
and observe the mentioned LR physics.

In the present paper, we push forward the previous theoretical analysis, focusing on a two-dimensional topological superconductor, realized on a square lattice loaded by spinless fermions and with an algebraically decaying pairing.

{\bf The model --} 
The Hamiltonian of our two-dimensional model is based on a square lattice  with $Lì^2$ sites and reads: 
\begin{align}\label{Ham}
H_{\mathrm{lat}} & = - w \sum_{\langle i,j \rangle} \left(c^\dagger_i c_{j} + \mathrm{h.c.}\right)  - \mu \sum_{r=1}^{L^2} \left(n_r - \frac{1}{2}\right) +
\\ \nonumber
&+ \frac{\Delta}{2} \sum_{s_x , s_y = 1}^{L} \, \left(\sum_{\ell=1}^{L-1}  \frac{c_s c_{s+\ell \, \hat{x}}}{d_\ell^{\alpha}}  + i \, \sum_{\ell=1}^{L-1}  \frac{c_{s} c_{s + \ell \, \hat{y}}}{d_\ell^{\alpha}}  \, \right)  + \mathrm{h. c.} 
\end{align}
 where $s = (s_x , s_y)$.
Working on a closed lattice,  we defined in \eqref{Ham} $d_\ell = \ell$ ($d_\ell = L-\ell$) for 
$\ell < L/2$ ($\ell > L/2$) and 
we choose anti-periodic boundary conditions \cite{nostro}.  We set $\Delta=2 w =1$, 
which does not qualitatively limit our analysis.

The spectrum of excitations of $H_{\mathrm{lat}}$, obtained via a Bogoliubov transformation, is:
\begin{equation}
\lambda_{\alpha}(\b{k}) = \sqrt{\left(\mu + \cos{k_x} + \cos{k_y}  \right)^2 + f_{\alpha}^2 (k_x) +  f_{\alpha}^2 (k_y)} \, ,
\label{eigenv}
\end{equation}
where $k_{x/y}= 2\pi \left(n_{x/y} +  1/2\right)/L$, 
$0 \leq n_{x/y}< L$, and
$f_{\alpha} (k) \equiv \sum_{l=1}^{L-1} \sin(k l)/d_\ell^\alpha$ (expressed in the thermodynamic limit as 
combinations of polylogarithmic functions \cite{nostro}). 

The superfluid ground-state of \eqref{Ham} reads:
\beq
\ket{\psi_{\rm gs}}_{\alpha} = \prod_{\b{k}} \Big(\mathrm{cos} \, \theta_{\alpha , \b{k}} + e^{i \, \phi_{\alpha , \b{k}}} \, \mathrm{sin} 	\,  \theta_{\alpha , \b{k}} \, a^{\dagger}_{\b{k}} a^{\dagger}_{-\b{k}} \Big) \, \ket{0} \, ,
\label{GS}
\eeq
with  $\phi_{\alpha , \b{k}} = \mathrm{arctan} \big(f_{\alpha}(k_y) / f_{\alpha}(k_x)$\big) and $\theta_{\alpha , \b{k}}$ defined similarly as for $p$-wave superfluids \cite{annett}.

If $\alpha \leq 1$, $f_{\alpha} (k)$ gives rise to 
singularities, at $k_x= 0$ and $k_y = 0$, in the spectrum;
the corresponding eigenstates, inheriting from these singularities some branch-cuts in their wavefunctions,  are notably
far from the minimum of the energy spectrum and they are dubbed "singular states" (SST) \cite{paper1,lepori2016mech}. 

{\color{black} Just considering the symmetry content, the model in \eqref{Ham} belongs to the $D$ class of the ten-fold way classification \cite{zirnbauer1997,ludwig2008}, since time-reversal is explicitly broken by the spinless nature of the fermions, while charge conjugation ${\mathcal C}$ is preserved: $H_{\mathrm{lat}} (\b{k}) = - U_c \, H^*_{\mathrm{lat}} (- \b{k}) \, U_c^{\dagger}$, with $U_c = \sigma_1$ and $U_c^2 = {\bf 1}$ \cite{ludwig2009}. Importantly, the  singularities at $\alpha \leq 1$ (say in $\b{k}_0$) do not change the definition of ${\mathcal C}$ above, since $U_c = \sigma_1$ both at $\b{k}_0 + \b{\epsilon}$ and $\b{k}_0 - \b{\epsilon}$, $\b{\epsilon} \to 0$.}

A relevant generalization of the model  \eqref{Ham} involves a LR hopping.
However, as explained in \cite{lepori2016}, this term can modify the structure of the SR phase diagram \cite{paperdouble,delgado2015}, but cannot
induce any new phase.
This happens because no branch-cut is produced by the SST (see below). 
We stress that this feature depends on the precise 
geometry of the chosen lattice: for instance in \cite{bettles2017} some LR effects are found from LR hopping on triangle and honeycomb lattices (loaded by hard-core bosons and with emerging fermionic behaviour).

{\bf Phase diagram --}
The spectrum in \eqref{eigenv} displays a critical line at $\mu = 2$ 
for every $\alpha>0$ and two critical lines at $\mu = 0$ and $\mu = -2$ for $\alpha >1$, there some continuous quantum phase transitions (QPT)s
occur. The phase diagram can be characterized further by the Chern number \cite{niu1985,bernevigbook}:
\beq
n = \frac{1}{2 \, \pi} \, \int_{\mathrm{BZ}} \mathrm{d} \b{k} \, \cdot \, \Big( \nabla_{\b{k}}  \, \times \,  \langle u_{\alpha, \b{k}} | \nabla_{\b{k}}  | u_{\alpha, \b{k}} \rangle  \Big)\, ,
\label{berry}
\eeq
where $|u_{\alpha, \b{k}} \rangle$ is the eigenstate corresponding to $\lambda_{\alpha}(\b{k}) $.  
This quantity vanishes if $|u_{\alpha, \b{k}} \rangle$ is continuously defined on the Brillouin zone \cite{bernevigbook}, while $n \neq 0$ requires some branch-cut on it. These singularities are encoded in the phases $e^{i \, \phi_{\alpha , \b{k}}}$ in  \eqref{GS}, and at $\alpha \leq 1$ are generated \emph{also} by the SST.

\begin{figure}[t]
\centering
\includegraphics[scale=1.2]{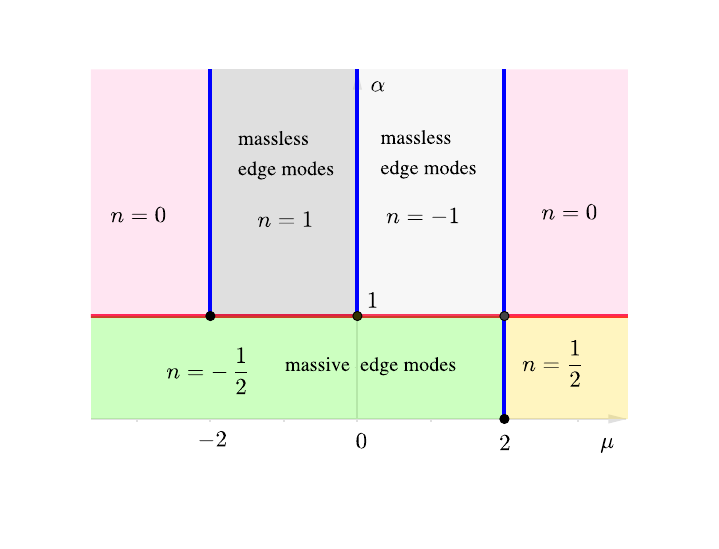}  
\caption{Phase diagram  at $T=0$ for $\Delta=2w=1$.}
\label{fig:phasediag}
\end{figure}

For $\alpha >1$ we obtain (see Fig. \ref{fig:phasediag}): \emph{i)} $n= 0$ if $|\mu|>2$, 
\emph{ii)} $n= -1$ if $0<\mu<2$, \emph{iii)} $n= 1$ if $-2<\mu<0$. Instead, if $\alpha<1$:
\emph{iv)} $n= -1/2$ if $\mu<2$ and \emph{v)} $n = 1/2$ if $\mu>2$.\\
For $\alpha>1$ one recovers qualitatively the two dimensional chiral lattice model for $p-$wave 
superconductivity, holding in the SR limit \cite{bernevigbook}.

Instead, the semi-integer numbers below $\alpha = 1$ do not make sense formally, since they cannot be associated to any homotopy class \cite{nakahara,volovik,ludwig2009} for the topology-inducing maps $\b{k} \to H_{\mathrm{lat}}(\b{k})$ (indeed discontinuous, due to the SST). {\color{black} Yet, 
they imply the emergence of two new (dubbed LR) phases, not continuously connected 
with the phases in the SR 
regime, and not belonging to {\color{black} the standard classification of topological insulators/superconductors} (where $n$ are integer) \cite{lepori2016}, despite the fact that $H_{\mathrm{lat}}$ collocates in the $D$ class of the ten-fold way according to its symmetries.} The same indication 
comes from  the QPT on the line $\alpha = 1$ that, remarkably, 
occurs without any mass gap closure in the spectrum \eqref{eigenv} (and without a first-order behaviour, as checkable from the derivatives in $\alpha$ of the ground-state energy of $H_{\mathrm{lat}}$ in  \eqref{Ham} \cite{nostro}). 
Since the mass gap is always finite if $\alpha <1$ and $\mu <2$, this QPT
notably prevents an unreasonable crossover from disconnected phases with different $n =0, \pm 1$ in the SR limit (see Fig. \ref{fig:phasediag}).

{\bf Role of the SST at $\alpha \leq 1$ --}
Further insight on the QPT at $\alpha =1$ is achieved analyzing the fidelity susceptibility  
 \cite{YouPRE2007, ZanardiPRL2007,pezze2017},
defined  by $\chi_{\alpha} = - \frac{d^2 \mathcal{F}}{d \alpha^2}$, where 
\beq
 \mathcal{F} = {}_{+}\bra{ \psi_{\rm gs} } \psi_{\rm gs} \rangle_- = 
 \prod_{\b{k}} \chi_{\alpha,\b{k}}
\eeq
\big($\chi_{\alpha,\b{k}}$ obtainable directly from  \eqref{GS}\big) 
 is the fidelity between two ground-states of 
the Hamiltonian  (\ref{Ham}), for slightly different values of the parameter $\alpha$:
$\ket{\psi_{\rm gs}}_+  =  \ket{\psi_{\rm gs}}_{(\alpha + x)}$ 
and $\ket{\psi_{\rm gs}}_- = \ket{\psi_{\rm gs}}_{(\alpha - x)}$, $x \to 0$. 
The logarithm of $\chi_{\alpha}$ diverges at $\alpha =1$; moreover  this divergence is found to be again a consequence of the SST: indeed 
there $\chi_{\alpha,\b{k}} \neq 1$ only if  $k_x = 0$ and/or $k_y = 0$.

It is interesting to single out the contribution by the SST for the Chern number in \eqref{berry}. This task, {\color{black} difficult following e.g. \cite{suzuki2005},} can be achieved for instance computing $n$ within an effective theory (ET),  obtained by a renormalization group (RG) approach and describing the vicinity of the massless lines,  {\color{black}  similarly to \cite{bernevigbook} for $\alpha \to \infty$. 
For simplicity, we focus on the line $\mu = 2$.
We follow the strategy in \cite{paper1}, retaining, along the decimation procedure on $H_{\mathrm{lat}}(\b{k})$ inducing the RG flow \cite{huang}, only states close to the minimum of $\lambda_{\alpha}(\b{k})$, at $k_x = k_y = \pi$, and close to $k_x= 0$ and $k_y = 0$, where singularities develop at $\alpha \leq 1$. Indeed, these states only are relevant for the dynamics close to criticality \cite{paper1}. The resulting action is written, in terms of two Majorana fields $\psi_{\mathrm{M}} (\tau, \vec{r})$ and $\psi_{\mathrm{H}} (\tau, \vec{r})$ \cite{paper1}}, as a sum of two independent terms, corresponding (in order) with the two sets of states above:
\beq
S^{(\alpha<2)} = S_{\mathrm{M}} + S_{\mathrm{AN}}^{(\alpha<2)} \, ,
\eeq
being $S_{\mathrm{M}}$ the (Euclidean) Dirac action \cite{peskin}, and the conformal invariance breaking action
\beq
\begin{array}{c}
S_{\mathrm{AN}}^{(\alpha<2)} = 
\frac{1}{2} \, \int  \mathrm{d} \vec{r}  \, \mathrm{d}\tau \, \bar{\psi}_{\mathrm{H}} (\tau, \vec{r}) \, K_1(b, M_1)\,  \psi_{\mathrm{H}} (\tau, \vec{r})  + \\
{} \\
+ \frac{1}{2} \,  \int  \mathrm{d} \vec{r}  \, \mathrm{d}\tau \, \bar{\psi}_{\mathrm{H}} (\tau, \vec{r}) \, K_2(b, M_2) \,  \psi_{\mathrm{H}} (\tau, \vec{r}) 
+ x \,  \leftrightarrow y \, ,
\label{ET}
\end{array}
\eeq
with $b = \alpha -1$, $K_1(b, M_1) = \Big[\gamma_{\tau} \, \partial_\tau \,  + \, c_{1x} \, \gamma_x \,  \partial_x^{b} + \, c_{1y} \, \gamma_y \,\partial_y^{b}  \, + \,M_1  \Big] $,  $K_2(b, M_2) = \Big[\gamma_{\tau} \, \partial_\tau \,  + \, c_{2x} \, \gamma_x \,  \partial_x + \, c_{2y} \, \gamma_y \,\partial_y^{b}  \, + \,M_2  \Big] $ ($c_{1x}$ and  $c_{1y}$ of the same order and $ c_{2x} \to 0$ along the RG flow),  $\{\psi_{\mathrm{M}} (\tau, \vec{r}) , \psi_{\mathrm{H}} (\tau, \vec{r}) \} =0$, $\gamma_{\tau} = -\sigma_3$, $\gamma_x =   \sigma_1$, and $\gamma_y =  \sigma_2$.  The symbols $M_{\rho}$, $\rho = \{1,2\} $, denote two renormalized masses: $M_{\rho} \to \infty$ if $\alpha >1$ along the RG flow, while
$M_{\rho} \to 0$ if $\alpha <1$.

Starting from $S^{(\alpha<2)}$,  the calculation of the Chern number $n= n_{\mathrm{M}} +  n_{\mathrm{AN}}$ proceeds as follows: 
the Hamiltonian matrix corresponding to $S_{\mathrm{AN}}^{(\alpha<2)}$ can be written in momentum space as
$H(\b{p}) = \hat{d}(\b{p}) \cdot \b{\sigma}$.
This gives a contribution \cite{bernevigbook}:
\beq
n_{\mathrm{AN}} = \frac{1}{8 \, \pi}  \int_{- \infty}^{\infty} \mathrm{d} \b{p} \, \epsilon_{abc} \epsilon_{ij} \,  d_a (\b{p}) \, \partial_{p_i}  
d_b (\b{p}) \, \partial_{p_j} d_c (\b{p})  \, ,
\label{berryeff}
\eeq
with $(i,j) = {\{x,y\}}$, and $\{a,b,c\} = 1-3$. One finds 
$n_{\mathrm{AN}}=0$  if $1<\alpha<2$, while, for  $\alpha<1$,   $n_{\mathrm{AN}} = \frac{1}{2}$
(due to the first term  only in \eqref{ET}, since $c_{2x} \to 0$  and $\partial_x$ does not produce any additional discontinuity), which adds to  
$n_{\mathrm{M}} =-1$ from $S_{\mathrm{M}}$ if $\mu<2$ ($n_{\mathrm{M}} =0$ if $\mu >2$) \cite{bernevigbook}, yielding $n  = - \frac{1}{2}$ \big($n=  \frac{1}{2}$\big), accordingly with the lattice results.
Importantly, the different values for  $n_{\mathrm{AN}}$ above and below $\alpha = 1$ depend on the different behaviours of $M_{\rho}$ along the RG flow, these masses describe indeed the QPT on this line \cite{paper1}. 

{\color{black} The stability of the LR phases against local disorder (e. g. in the 
chemical potential term $\propto \mu$) is inferred following \cite{rodriguez2003}. There, using RG arguments,  the SST are shown to be not localized by disorder (differently from
the other eigenstates) below a critical $\alpha_c$, in our case  $\alpha_c = 2$.}

{\bf Asymptotic decay of two-points correlations --}
The appearance of the QPT at $\alpha = 1$ and of the LR phases below this threshold requires LR correlations \cite{lepori2016}. The same feature justifies the evolution of the edge modes described in the following.
The two-point correlations are computed from the Bogoliubov transformations diagonalizing the Hamiltonian  \eqref{Ham} \cite{refpar,paperdouble}.
\begin{figure}
\centering
\includegraphics[scale=0.26,angle=0]{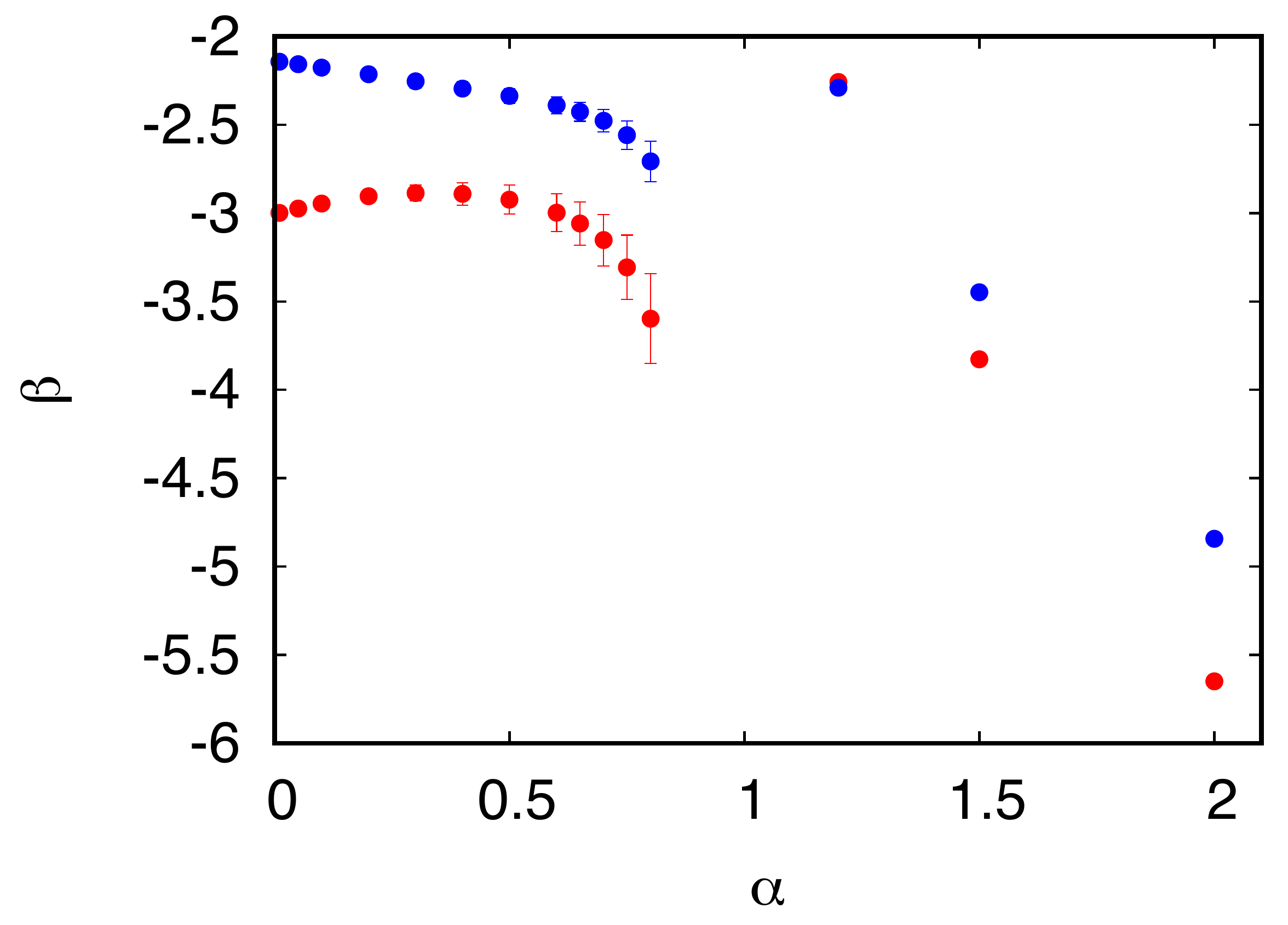}
\caption{Decay exponents $\beta$ of  $\langle c_i^{\dagger} c_j \rangle$ (red) and  $\langle c_i^{\dagger} c^{\dagger}_j \rangle$ (blue), for $L = 1500$,  $\mu = 0.5$  and $\alpha$ varying.  The regime close to $\alpha = 1$ is not plotted, due to the presence of large finite-size effects.}
\label{fig:corr}
\end{figure}
Both correlations asymptotically display an algebraic decay, also if the mass gap is nonvanishing. In  Fig. \ref{fig:corr} we report  their decay exponents $\beta$ at fixed $L = 1500$ and  $\alpha$ varying. Similarly to \cite{nostro}, we  encounter large errors around $\alpha = 1$, likely due to the QPT. 
Compared to the SR regime, the decay becomes much slower at 
$\alpha < 1$ (and essentially for every $\mu$). 
Notably, even there, $\beta$ remains smaller than 2, implying that  multipartite entanglement cannot be detected using local operators involving (linear combinations of) $c_i$ and $c_i^{\dagger}$. This results enlarges the picture  achieved for the LR Kitaev chain \cite{pezze2017,hauke2015}.

{\bf Bipartite entanglement --} 
\begin{figure}
\begin{center}
\includegraphics[scale=0.25]{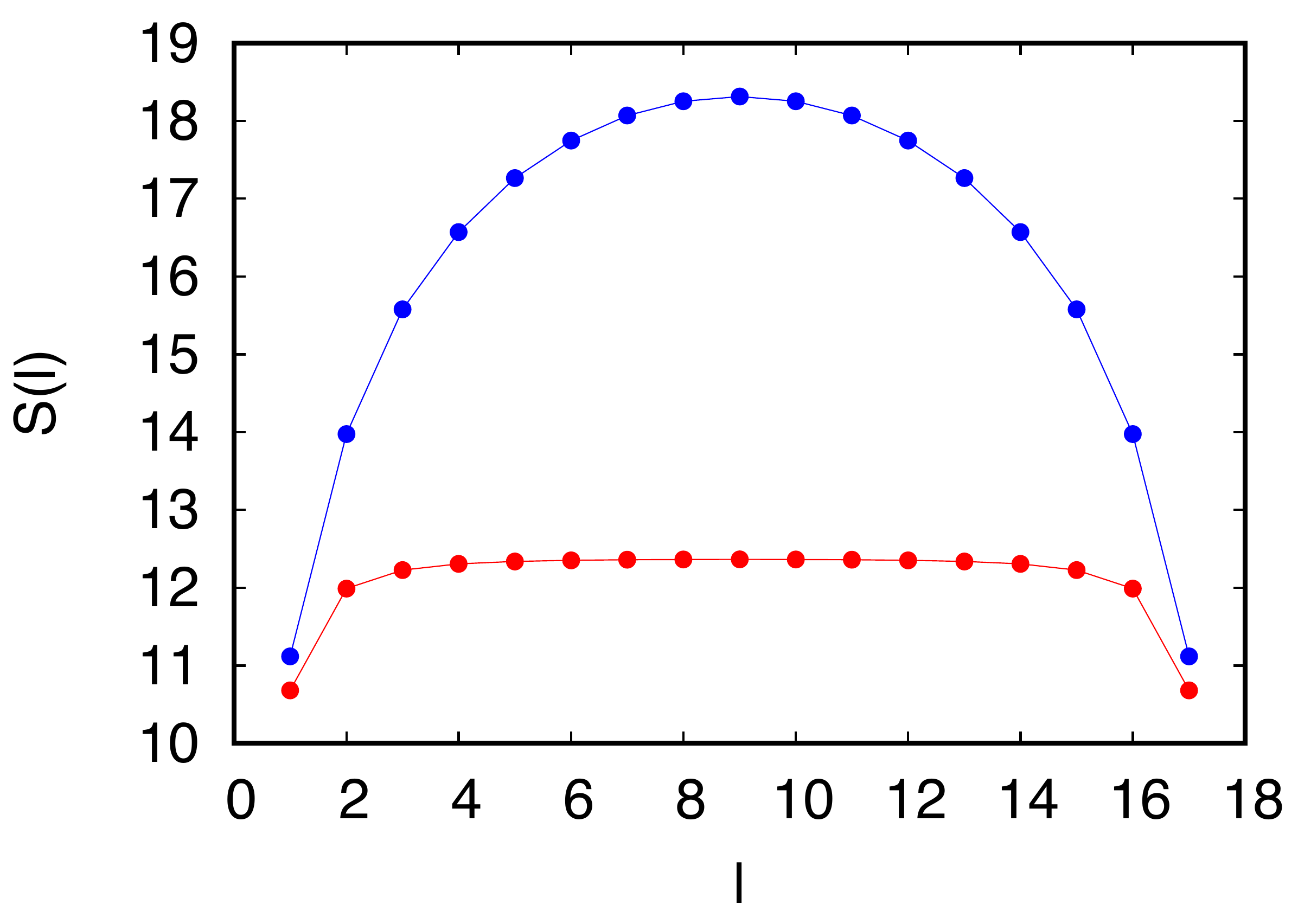}  
\includegraphics[scale=0.26,angle=0]{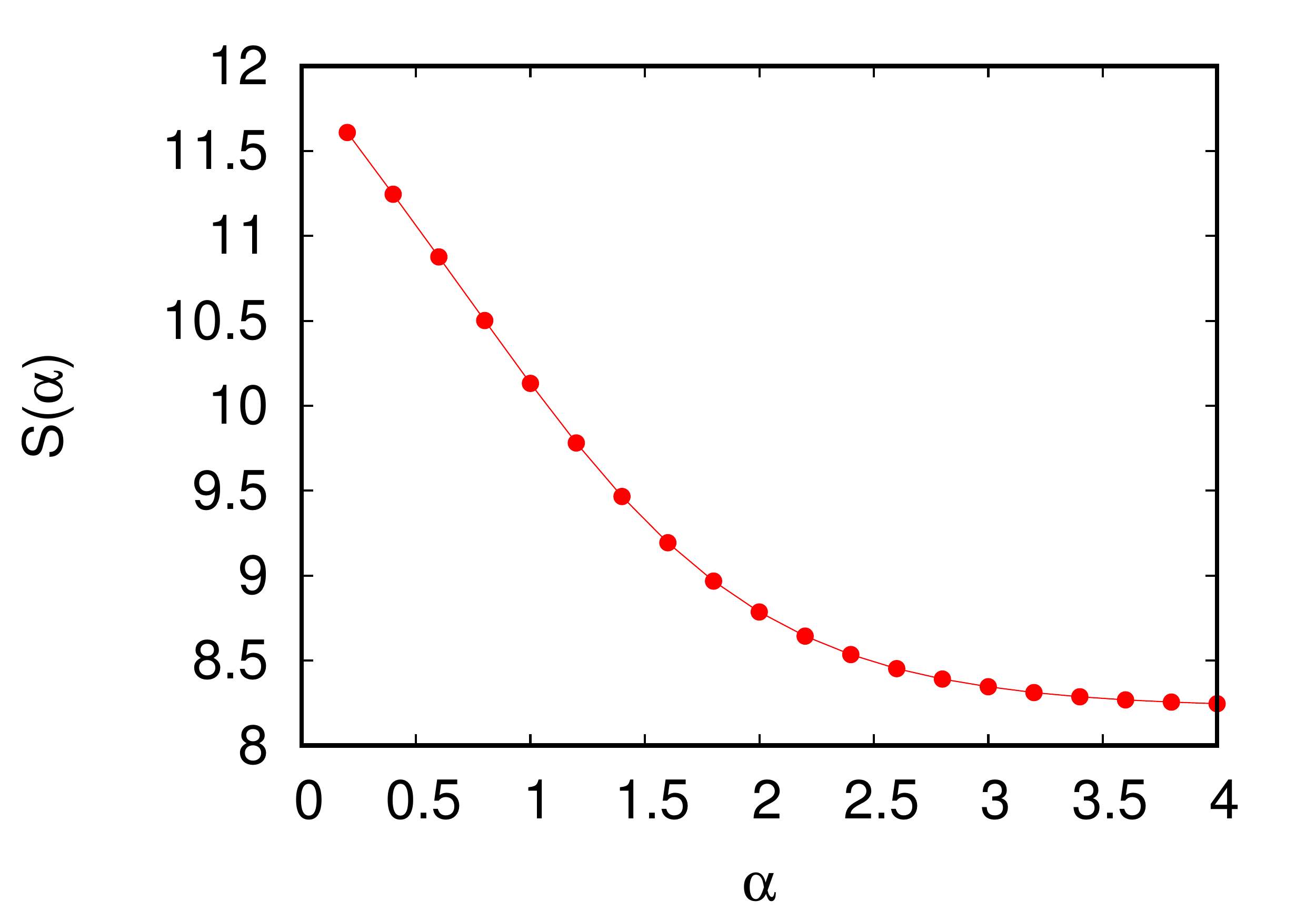}
\end{center}
\caption{Behaviour of the Von Neumann entropy $S$ for $L = 18$ and $\mu = 0.5$.
(Up)   $S(l)$ for $\alpha = 10$ (red) 
and for $\alpha = 0.5$ (blue). Notice the difference in the two cases for the variation  around $l= \frac{L}{2}$.
 (Down) Same quantity for fixed $l = \frac{L}{2}$ and $\alpha$ varying.}
\label{fig3}
\end{figure}
Further information on the LR phases can be achieved by the analysis of the VNE $S$ between two sublattices, for varying  $\mu$ and $\alpha$. We adopt  the technique in \cite{peschel2011}, exploiting
the correlations $\langle c_i^{\dagger} c_j \rangle$ and $\langle c_i^{\dagger} c^{\dagger}_j \rangle$: we define the matrix 
$M_{ij} = \langle (a_{2 i -1} , a_{2 i}) (a_{2 j -1} ,  a_{2 j}) \rangle$,
where $a_{2 l -1} = c_l + c_l^{\dagger}$ and $a_{2 l} = i \, (c_l - c_l^{\dagger})$ are Majorana operators, the indices $i , j = 1, \dots , s$   running on the sites of the subsystem A with $s$ sites.
This matrix has $s$ pairs of eigenvalues $1 \pm v_n$, from them the VNE is straightforwardly expressed.  \\
In this way, we evaluate $S$ on an antiperiodic torus with $18^2$ sites, selecting subsystems 
A with $18 \, l$, $l= 1, \dots, \frac{L}{2}$ sites, obtained cutting the torus twice along one direction. 
In this framework, one recovers the area law for $S(l)$ in that it becomes independent of $l$, $S(l)  \approx S_0$, for 
$l$ sufficiently 
far from $l =0$.
By direct inspection, we find that this picture is realized for every $\mu \neq 0 , \pm \,  2$  at 
$\alpha \gtrsim 1$ (finite-size effects are found  to be relevant close to $\alpha = 1$). On the contrary, 
below this  approximate threshold, $S(l)$ is no longer constant, signaling an area-law violation. The coincidence 
between this violation and the divergence of the quasiparticle spectrum has been also suggested in \cite{ares}.
The two situations are shown in Fig. \ref{fig3} (up), notice the difference in the variation for $S(l)$ around $l = \frac{L}{2}$.
Even if the small dimensions of the analyzed square lattices forbid to quantify the violation at $\alpha \lesssim 1$, in our examples this is found compatible with a logarithmic deviation. {\color{black} The violation of the area-law is strongly required by the appearance of the LR phases:  the continuous transition at $\alpha = 1$ needs a diverging correlation length, due, in the absence of a vanishing mass gap, to the LR coupling. The same divergence is testified by the algebraic tails of correlations and induces the area-law violation, as in SR systems \cite{plenio2010,lepori2016}.}
Finally, a peculiar behaviour is also observed at small $\alpha$ for the half-square VNE  $S(\alpha)$, reported in Fig \ref{fig3} (down): around this line $S(\alpha)$ starts to increase much more rapidly than at higher $\alpha$. This behaviour becomes more pronounced and picked around $\alpha = 1$ as $L$ increases. 

{\bf Edge mode structure  --} 
\begin{figure}
\centering 
\includegraphics[scale=0.30,angle=0]{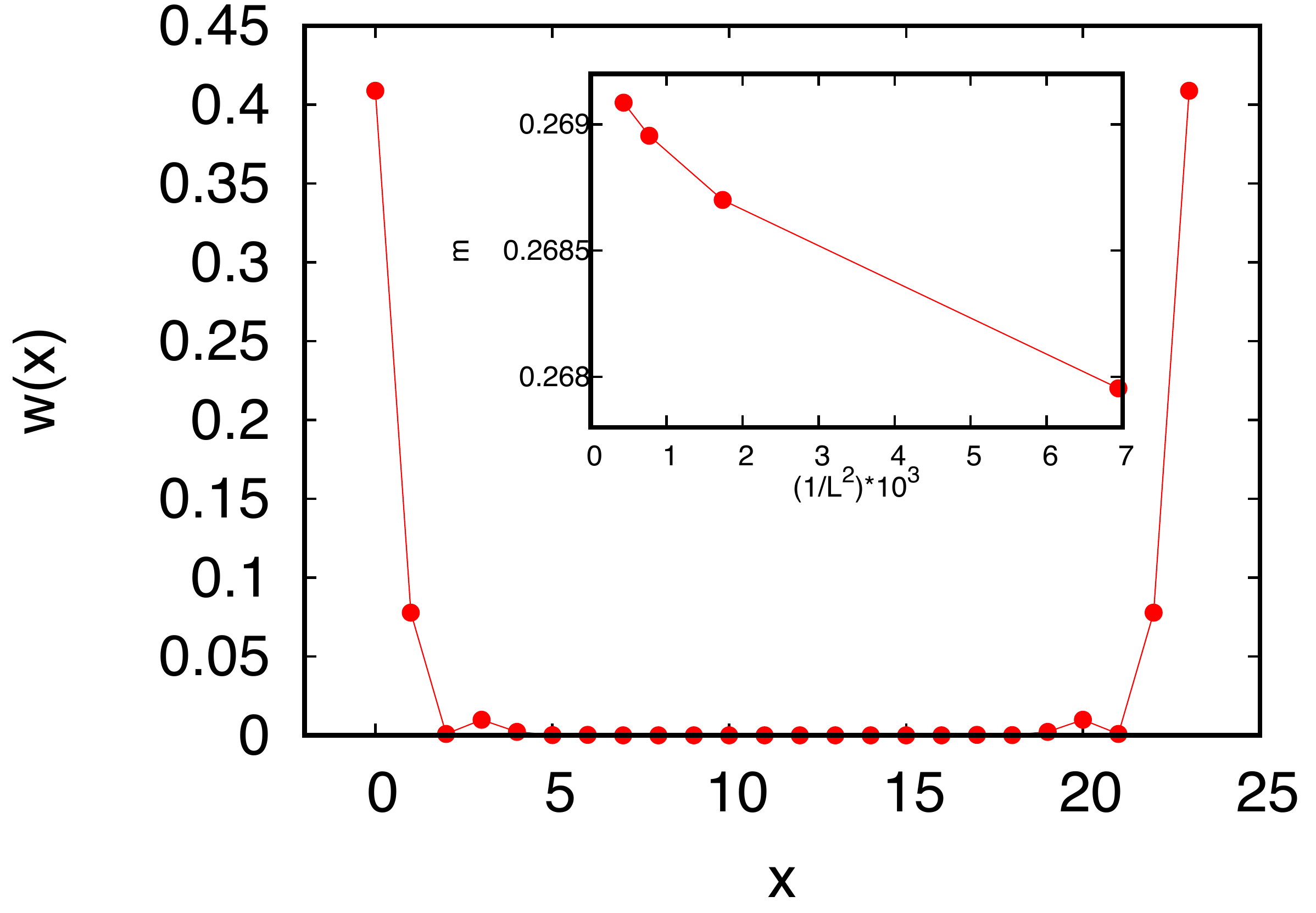}
\caption{(Main panel) Square of the wavefunction, $w(x)$, of the massive edge modes at $L = 24$, $\mu = 0.5$ and $\alpha = 0.5$, for $y = \frac{L}{2}$ and  varying $x$.
(Inset)  finite-size scaling of the mass for the edge modes at $\mu = 0.5$ and $\alpha = 0.5$, for $L=12-48$ varied in steps of 12. A nonvanishing mass in the thermodynamic limit is obtained.}
\label{fig2}
\end{figure}
The peculiar values for $n$ at $\alpha<1$ induce to investigate the edge mode structure of the Hamiltonian \eqref{Ham}. 
Therefore we 
numerically computed its spectrum, assuming both open boundaries and a compactified  direction (again with 
antiperiodic conditions).
 For $|\mu| < 2$,  massless edge modes appear at $\alpha \gtrsim 1$ when $n \neq 0$,  similarly to the SR limit \cite{bernevigbook}. This is an example of the so-called bulk-boundary correspondence  \cite{zirnbauer1996,zirnbauer1997,ludwig2008,kitaev2009,ludwig2009,ludwig2010}. The edge modes correspond in the mixed model to zero-energy (in the thermodynamic limit) eigenstates with the (square of the) wavefunction  picked around the edges,  and well separated in energy from the higher-energy states, instead spread in the bulk. This situation is qualitatively equal to the one-dimensional open  (LR) Kitaev chain \cite{nostro}. For the open model, edge modes manifest as a continuum of states with linear dispersion crossing the zero-energy line and linking the particle and hole bands \cite{hasankane}.\\
On the contrary, at $\alpha \lesssim 1$ edge modes with a nonvanishing mass
 appear at $\mu <2$ (while no edge mode occurs at $\mu >2$, although there $n \neq 0$): in Fig. \ref{fig2} (inset) a finite-size scaling plot of the edge mass is reported, showing that it does not vanish even in the thermodynamic limit.  The massive edge modes (MEMs) arise in the mixed model as
 eigenstates with a nonzero energy, but still well separated in energy from the higher-energy ones; the square of their wavefunction  results again picked around the edges (see Fig. \ref{fig2} (main panel)). At least in this situation, the appearance of the MEMs can be explained by the failure of the edge operators classification in \cite{turner2011,lepori2016}, due to large LR correlations between bulk and edges; {\color{black} related to this fact, the multiplets of the Schmidt eigenvalues are not constrained to have even multiplicity, as instead at $\alpha >1$.}
In the open model, the MEMs do not cross the zero-energy line anymore, so that an energy gap appears
around this line; since this gap refers to the ground-state energy, its presence implies the absence of single-fermion edge conductivity, instead present in the SR limit.\\
The stability of the MEMs against a local disorder parallels that of the LR phases discussed above. The possibility of a hybridization mechanism (of the massless edge states at $\alpha \gtrsim 1$) generating the MEMs, as in one dimension \cite{paperdouble,delgado2015,neupert2016}, remains a likely but open issue in the mixed case, while an edge reconstruction mechanism \cite{capone2017} seems ruled out by the change of $n$ passing through $\alpha = 1$.

{\color{black} {\bf Relation with LR bulk topology -- } The appearance of the MEM can be put in direct relation with the bulk properties.
For this task, when $\alpha<1$ we perform the integration in  \eqref{berry}  avoiding the singular lines $k_x = 0$ and $k_y = 0$; this effectively 
restores the continuity of the maps $\b{k} \to H_{\mathrm{lat}}(\b{k})$, that therefore can still define a topology, as in the SR limit. This procedure, {\color{black} automatically performed following \cite{suzuki2005},} yields
  a shifted Chern number $\tilde{n} = n - n_{\mathrm{AN}} $ ($n_{\mathrm{AN}} = 1/2$ around $\mu =2$, as in \eqref{berryeff}): in this way, at $\mu<2$, where a MEM is present,  we obtain correspondingly $\tilde{n} =-1$, while at $\mu >2$, where MEM are absent, $\tilde{n} = 0$. 
Therefore, if  $\alpha<1$ it is still possible to envisage the notions of topology and bulk-boundary correspondence. The same procedure induces to classify the LR phases {\color{black} both by the singularities of the map from $H_{\mathrm{lat}}(\b{k})$, as partially described in \cite{lepori2016}, and, once they are removed as above, according to the usual scheme for SR systems \cite{zirnbauer1996,zirnbauer1997,ludwig2008,kitaev2009,ludwig2009,ludwig2010}.}

{\bf Conclusions  --}
The fermionic lattice studied in this paper is the first two-dimensional one where qualitative deviations from the classification of topological insulators/superconductors, due to LR Hamiltonian terms, are
described: there new phases occur, induced by extreme LR correlations (resulting e.g. in the area-law violation for the Von Neumann entropy).  
A notion of bulk topology can be still defined for the LR phases, related with a nontrivial massive edge structure (in turn implying the absence of edge conductivity) and still characterized by  bulk invariants, which are integer if suitably redefined to trace out the singularities from the LR couplings. 

Our results, confirming and enlarging the picture drawn in \cite{nostro,paperdouble,paper1,delgado2015,lepori2016} mainly for one dimensional LR models,   are a valuable starting point for the full classification of the (topological) LR phases for insulators/superconductors. Moreover, they allow for the comprehension of experimentally achievable LR models, e.g. in dissipative systems, as the recent proposal \cite{bettles2017}, and in cavities \cite{revritsch}. In \cite{bettles2017} the absence of massive edge modes can be a finite-size effect, forbidding to probe the edge hybridization (crossing at high enough values for $\alpha$) momentum.
Finally, from a more general perspective, our study highlights the effects of extreme LR correlations in many-body systems.} \\

{\bf Note added -- } Just before the conclusion of the present manuscript, a paper appeared \cite{delgado2017}, dealing with a model similar  to ours. The results contained there, especially about the edge structure, appear consistent with ours.\\

{\bf Acknowledgements --}
The authors are pleased to thank  P. Calabrese, M. Capone, S. Ciuchi, M. Giampaolo, M. Mannarelli,
 I. Peschel,  L. Pezzé, A. Smerzi, A. Trombettoni, and D. Vodola for the enlightening discussions and correspondence. 
 S. P. acknowledges support by a Rita Levi-Montalcini fellowship of the Italian Ministry of Public Education. Part of this work has been
performed during the  workshop "From Static to Dynamical Gauge Fields with Ultracold Atoms",
in the Galileo Galilei Institute for Theoretical Physics, Firenze,  May 22th - June 23th 2017.

\end{document}